# Perception of Motion and Architectural Form: Computational Relationships between Optical Flow and Perspective

Arash Sangari[1], Hasti Mirkia[2] and Amir H. Assadi[3]

[1] Department of Electrical and Computer Engineering, University of Wisconsin, USA
[2] Persepolis Research Group, University of Wisconsin, USA
[3] Department of Mathematics, University of Wisconsin, USA

**Abstract.** Perceptual geometry refers to the interdisciplinary research whose objectives focuses on study of geometry from the perspective of visual perception, and in turn, applies such geometric findings to the ecological study of vision. Perceptual geometry attempts to answer fundamental questions in perception of form and representation of space through synthesis of cognitive and biological theories of visual perception with geometric theories of the physical world. Perception of form, space and motion are among fundamental problems in vision science. In cognitive and computational models of human perception, the theories for modeling motion are treated separately from models for perception of form.

## 1   Introduction

Modeling visual perception of form and motion has been studied extensively with many success stories. Primate brain's perception of motion from optical flow is attributed to dynamics in networks of the visual areas V1 through MSTd [5]. In contrast, cortical processes required for perception of spatial form (e.g. as in the interiors of buildings) are still largely unknown. Computational models and theories of motion perception have not been studied in the context of perception of architectural forms; that is, shape perception and formation of the percept of planes in perspective are studied separately.

Below, we provide the theory and a model for perception of interior space in architecture. We propose the existence of neuroanatomical loci that play a pivotal role in perception of perspective, and demonstrate a mechanism for visual perception of interior spaces based on eye movement. We conclude that learning to estimate motion from optical flow is similar to learning to estimate a single vanishing point. Both perceptual tasks, therefore, potentially employ the same local circuit in the MSTd. To enhance our theory, we show that an ANN with sufficient accuracy in predicting motion from optical flow learns the tasks of detecting vanishing points more efficiently. In contrast, training a generic ANN with random initialization will not reach the corresponding level of accuracy under the same protocols. Our simulation suggests a neurobiological hypothesis for "sharing of the neuronal pathways" for detection of vanishing points and estimation of direction of motion from optical flow. More precisely, the flow of information from the retina through primary and higher areas of the human visual cortex employs plasticity of the cortical circuits in MSTd that are known to compute motion from optical flow. The "MSTd Hypothesis for Perception of Perspective" proposes that the vanishing point detecting neurons are in MSTd.

In view of the emerging advances due to understanding the role of eye movement in mental health research, we plan to develop a predictive theory for human affective response to variation of spatial features of interior spaces. This article provides the computational model and tools for such a theory.

## 1.1 Geometry Perception

The study of 3D perception from 2D images is the main issue of vision science in both biological and computer systems. Depending on the projection method used for mapping 3D space to a 2D plane, different images may be resulted. Among these projection methods, perspective projection is known as an accurate mapping for reconstructing geometrical features of real objects. One important concept in perspective projection is the "vanishing point". The vanishing point of a straight line under perspective projection is that point in the image beyond which the projection of the straight line cannot extend. Therefore, the vanishing points correspond to the "points at infinity" in the receding direction of the parallel straight lines in space.

Computationally, the vanishing point in interior spaces with a single vanishing point can be found by finding the point where most of the salient straight lines in the image intersect with each other. The salient straight lines can be found using processes such as edge detection and line extraction. The mechanism of edge detection in the brain is related to the lateral inhibition in the receptive field of the visual sensory system [6]. A sample of how edge detection occurs in an interior space is shown in figure 1-b.

The Gestalt principle of continuity shows how the brain can fill in the gaps between the dots in the image and extract the major lines. The Hough transform can be used to extract the slope and the constant term of lines in the image, based on black and white result of edge detection algorithms [2]. The line drawings made from the interior spaces of figure 1-b carry pieces of information that the brain needs to estimate the orientation of surfaces and compute the perception of the interior space in perspective. By extending these lines and finding the intersection points, we can computationally find the vanishing point (figure 1-c).

In real word experience, additional information such as texture, color and cues provide much more useful information for richer perceptual experience of interior spaces [1, 3]. However, this paper focuses on simpler geometric objects as illustrated in the line drawings below.

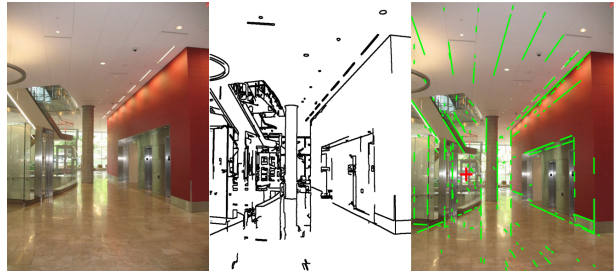

**Fig. 1.** Two steps in detection of vanishing point: edge detection of local circuits in area V1 and the thalamic feedback system provide the information about the lines as above. We have simulated the extraction of lines using Hough transform. (a) Left: an image from the Wisconsin Institute for Discovery (WID). (b) Center: binary output of edge detection. (c) Right: overlay of the Hough transform output onto the original image.

**1.2 Motion Field Estimation**

Structure determination algorithms such as shape from shading are often used to model perception of 3D forms from 2D images. However, their accuracy is correlated to the complexity of the image. As it is known in neuroscience, the brain takes advantage of motion in the image plane to estimate 3D form. Similarly, additional temporal dimension could be used to design algorithms with increased accuracy in modeling 3D shape perception.

The relative motion between image plane and scene can be estimated from successive image frame gradients in time and space [6]. If the sequence of image frames is denoted by a 3D array $I(x,y,t)$, the projection of the relative velocity vector onto the image plane is:

$$\frac{\partial I}{\partial x}V_x + \frac{\partial I}{\partial y}V_y + \frac{\partial I}{\partial t} = 0 \tag{1}$$

This equation for each pixel of the image in each frame yields a pair of relative velocity components:

$$V(m,n) = \left(V_x(m,n), V_y(m,n)\right) \tag{2}$$

The latter pairs form a complex-valued matrix as in:

$$V(m,n) = V_x(m,n) + iV_y(m,n), \quad i^2 = -1 \tag{3}$$

Since this matrix shows the flow of image data (in pixels) motion in an image plane, in computer vision and vision science this matrix usually is called an optical flow matrix.

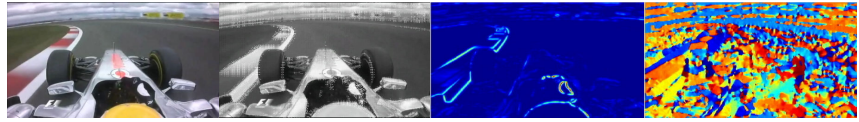

**Fig. 2.** This figure shows estimation of motion field (optical flow) from a movie. Far left: Lewis Hamilton in his Nürburgring 2011 qualifying lap. Next, gray scale image of the same frame with optical flow vectors. Third from left: heat map showing the magnitude of optical flow vectors (blue is zero and red is the largest magnitude). Right: The phase heat map.

### 1.3 Eye Movement

Our long-term objective is to develop a unified dynamic framework for modeling perception of geometric forms and motions. The key biological observation is the dynamic nature of vision. Saccades are major players in visual perception of the physical world [4]. Further, cortical neurons have receptive fields ([4] pp. 41-43), roughly, a cone-shaped region of space measured in terms of the visual angle. It is believed that the relatively limited light rays passing through that cone stimulate neurons to eventually invoke mechanisms of visual attention. The concepts of the receptive field and visual attention lead to a number of significant hypotheses concerning neuronal mechanisms for perception of regularity in the viewed images. Among them, one could hypothesize that there is a biologically realistic computational model of '*intermediate-level vision*' that elucidates the process of neuronal detection of visual regularity in the presence of repetitive light patterns on the retina. This mechanism is a hybrid of hierarchical receptor field theory and visual attention, in the form of a network with multiple feedbacks.

Accordingly, a series of spike trains that encode a temporal sequence of short intervals $T_j$ preprocess the visual stimuli. The collection of $T_j$ undergo computational loops that implement an adaptive process of forming correlations among $T_j$ by improving the length of the interval $T_j$ and the choices of the subsequences (in reference to saccadic movements that need not be in a uniform order). The results of such recursive process is an adaptively computed output, which is dominated by stable and coherent patterns of dynamics that form the inputs to the higher level visual processing centers. Repetition of highly correlated spike trains result in arrival of even more highly correlated spike trains to the higher level visual centers, which in terms of Bayesian models increase the likelihood of the recognition of that common pattern, thus bringing visual attention mechanisms to take over and pursue the process of perception, cognition and interpretation/action. One may well speculate that a sequence of saccades repeatedly brings to the fovea the implicit information in the scene that translates to repetition of highly correlated spike trains in specific intervals.. We refer to the mechanism thus formulated as the Adaptive Eye Movement Hypothesis (AEMH.)

Perceptual geometry investigates physiological evidence to establish the low-level computations of visual, tactile motor and auditory processes that contribute to the ability to perceive regularity in structure [1] [2]. For example, perceptual geometry predicts that the human perception of form and spatial organization of planes in architectural interior space could be modeled via the computational elements that perform the local computations for reconstruction of self-motion from optical flow. Eye movement is the key ingredient in models of vision that refer to AEMH. In the context of this paper, AEMH refers to the ubiquitous and continual presence of eye movements of several sorts (the tiny, almost instantaneous jitters of the eyeballs) that provide the impressions of frames of a movie, even when we fixate and seemingly try to maintain our eyes, head and neck motion-less. Suffice it to say that according to AMEH, the perception of natural surfaces, bounding planes and other geometric elements that contribute to perception of architectural form arises from integration of the flow of sensory processes and subsequent neuronal network information processing of a temporal sequence of "single frames."

It is certainly true that saccades and other eye movements capture images that could change orientation and direction of fovea; thus, how such discontinuous current of information from scene geometry is a puzzle that could give rise to a cohesive and continuous Gestalt of the surfaces delimiting the spatial geometry. In a nutshell, the processes in the neuronal network need not follow the temporal order of saccades or other eye movements. Rather, short-distance feedback loops, dynamics of local attractors in coordinated sub-collections of local circuits, and the phenomena that are collectively referred to as "visual attention" could maintain just enough records of the "seemingly single image frames," in such a way that the processing could sustain an orderly organization of the pieces of the percept.

Therefore, our first attempt focuses on modeling adaptive processes that simulate saccades (Latimer, 1996). Below, we consider a simplified computational model that detects the Gestalt of lines in the formation of vanishing points, as in perspective drawings. This model is robust with respect to noise, partial occlusion and some irregularity in the unavailability of directly observable edges. Moreover, it lends itself to generalization to visual attention and theories of active vision.

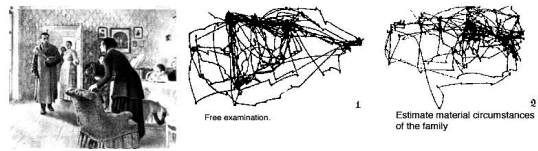

**Fig. 3.** Eye movement creates a sequence of snapshots (Alfred L, 1967 [7]). In our model the tree of eye movement creates a flow of visual information comparable to the frames of a movie.

## 2  Contribution to Value Creation

Advances in computational sciences and engineering have opened many possibilities for development of robotics and automation of routine tasks. For example, robot motion planning provides myriad possibilities for transportation and delivery robots inside hospitals and other public or private places. Meanwhile, much remains toward the development of robots with higher levels of learning or proper response to human needs. In this paper, we have demonstrated that perception of architectural form (a challenging problem for robots) could be developed with the same types of computations that are needed in estimation of motion from optical flow (a problem which has made great advances.) In a broader context, the follow-up of this research will also contribute to better understanding of the human response to the architectural form, thus enabling the architects and engineers to design habitats better suited for populations with specific needs or of specific genre of affective and cognitive abilities. In summary, this research offers many possibilities through applications to industrial and scietal value creation.

## 3   Method

Below, we outline a unified framework for perception of geometry and motion in interior spaces. We designed two 3D interior spaces in a virtual reality environment. We used a PCA for reducing the dimension of successive frames of this simulation and used some of the most informative components in training an ANN (*Optiflonet*) for estimation of direction of movement on an image plane. Then we demonstrated that this neural network with a slightly different configuration can extract vanishing point of still images.

### 3.1   Construction of simulation of training sets for neural network

In our interior model, we have simulated the sensory information by architectural rendering software: 3D Max. In order to test complex kinetics, virtual reality is used in rendering simple interior spaces (figure 4); we compared extraction of lines from simulated scenes versus extraction of lines in natural scenes.

### 3.2   Learning motion heading direction from optical flow

When an observer moves in an interior space, optical flow provides a robust mechanism to estimate the motion and heading direction. The optical flow vectors indicate the distribution of stimulated local circuits that detect motion and heading direction.

Optical flow vectors, (the arrows in the figure 5) can provide the input for training our neural network *Optiflonet*. Validation and testing of neural networks are performed using 100*100 arrays with sparse sampling of the optical flows. As illustrated in figure 5, for simple interior spaces, optical flow vector can be approximated by a sparse matrix. Therefore, we use a principal component analysis for extracting the most informative components of optical flow vectors in all frames.

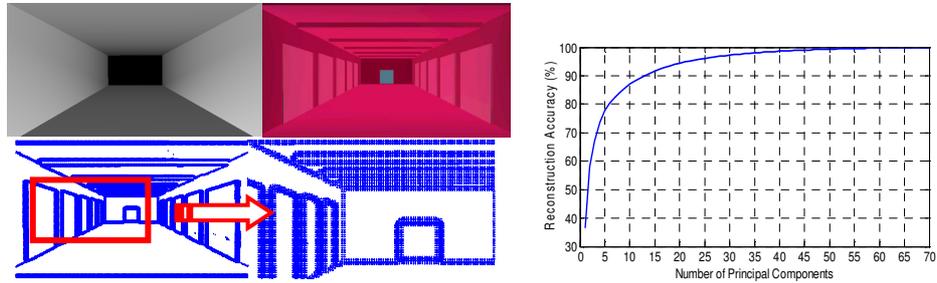

**Fig. 4.** (Left top) Examples for training, validating and testing an ANN to extract vanishing points. **Fig. 5.** (Left bottom) Optical flow vectors in simulated interior space movement. **Fig. 6.** (Right) Reconstruction accuracy vs. the number of principal components, e.g. 20 principal components has reconstruction accuracy of about 95%.

We use the first 20 principal components of image frames in order to reduce the required number neurons in our neural network while avoiding reconstruction errors (figure 6 shows the relationship between number of principal components and reconstruction error in our case).

Figure 7 shows the whole structure of the *Optiflonet* system to estimate direction of motion on an image plane. The architecture of neural network that we used is also presented in figure 8.

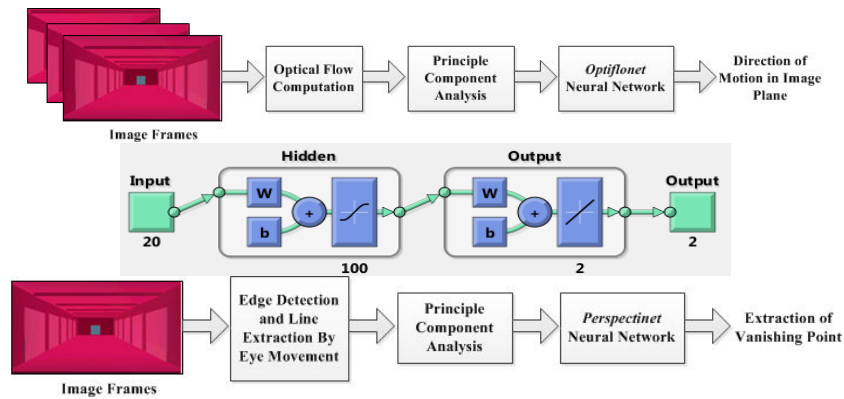

**Fig. 7.** (Top) Flowchart of *Optiflonet* to estimate direction of motion. **Fig. 8.** (Middle) Shows the architecture of *Optiflonet*, with two layers, and 100 neurons in the hidden layer. **Fig. 9.** (Bottom) Flowchart of *Perspectinet* for extraction of vanishing point.

To test our theory we designed a neural network that learns to extract vanishing points from perspectives called *Perspectinet*. This network incorporates a simplified model of eye movement, where the scene is sampled as a flow of consecutive snapshots of interior spaces. The subsequent layer of *Perspectinet* receives dynamic information from the scene.

Design of *Optiflonet* is based on the figure 9. The optical flow from movement on the retina provides the input to the next layers of visual sensation and visual perception of motion.

### 3.3 The proposed model for estimation of information from perspective

We trained *Optiflonet* using examples of motion in an interior space where each frame is a perspective projection of the interior space. The observer motion provides different projections on the retina, and consequently, differing perspectives with a possible shift of the vanishing point. The perspective could be tested in two ways. First, by turning the dynamic scene (perspective images) into movie that illustrates the effect of eye movement, the subsequent blocks will have layers of network and connection weights that are inherited from training the *Optiflonet*.

Our problem is how to introduce and measure the performance of the *Perspectinet* when test images are from interior space without any training on interior space forms. In other words, the local circuits of *Perspectinet* in the visual motion processing layers are used as local neural circuits that are employed for computations in movies of eye movement snapshots from interior spaces forms.

There is a second method for feasibility study of the "MSTd Hypothesis". Namely, we train two neural networks and compare their performance and accuracy. The first one is the *Perspectinet*, and the second ANN is a copy of *Perspectinet* whose initial synaptic weights are determined from training as local circuits of *Optiflonet*. These two ANNs are, now, trained, validated, and tested for accuracy using the exact same input data and the same protocols for stopping time. The test for validity of MSTd Hypothesis is the superior performance of the second ANN within the same training time and exposure to stimuli.

## 4 Results

In order to train *Optiflonet* we used back propagation method on our training data set. The training data set contained the 150 frames from a simulation of walking through a corridor like that in figure 4, right. We used another set of 150 frames for validating and testing of our neural network. Figure 10 shows the convergence of mean square error during the training, validating and testing.

Figure 11 shows the histogram of output errors in these procedures. Based on this histogram we can find that the norm of output error of *Optiflonet* is highly concentrated at about zero pixel errors in detection of motion in the frames. Figure 12 also shows the error in estimation of motion heading in the horizontal and vertical axes of each frame separately. Since in training and test data we used simulated walking, the kinetics of movement (velocity and orientation of camera) through interior space was the stochastic variable.

Figure 13 presents a time series of camera velocity and orientation during recording frame data during training. The orientation of the camera is shown as 3 components of the unit vector parallel to the camera orientation.

Finally, error in detection of location of the vanishing point in different frames is illustrated in two histograms in figure 14. The left histogram shows the vertical error while the right histogram shows the horizontal error. The maximum error along the horizontal axis is less than 2 degrees and less than about 13 degrees along the horizontal axis.

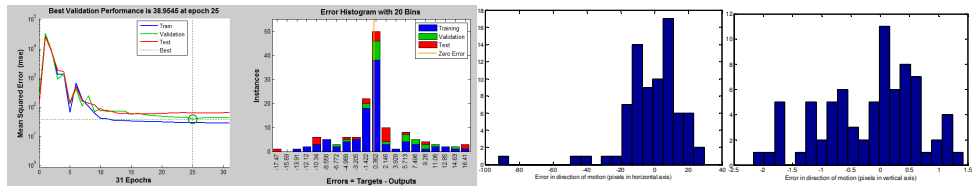

**(Fig. 10-12 Left to right.) Fig. 10.** Mean square error during the training, validating and testing of *Optiflonet* neural network. Best validation performance was achieved in the 25$^{th}$ epoch. **Fig. 11.** Histogram of output error during the training, validation and testing of the *Optiflonet* neural network. **Fig. 12.** Histograms of output error during the testing of the *Optiflonet* neural network. (Second right) error in estimation of movement heading along the horizontal axis. (Right) error in estimation of movement heading along the vertical axis. The unit of estimation error in these histograms is the pixel. Since the field of view of each frame includes 100 pixels covering 45 degrees, each pixel would cover a 0.45 degree of field of view.

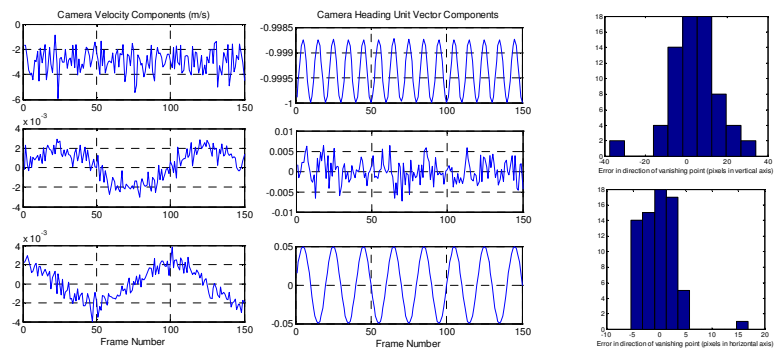

**Fig. 13.** (Left) Time series of camera movement (right plots) and orientation (left plots) during the recording data for neural network training. **Fig. 14.** (Right) Histogram of output error during the testing of the *Perspectinet* neural network. (Top) error in estimation of the vertical component of the vanishing point in the image. (Bottom) error in estimation of the horizontal component of the vanishing point in the image. The unit of estimation error in these histograms is the pixel. Since the field of view of each frame is comprised of 100 pixels, covering 45 degrees, each pixel would cover a 0.45 degree of field of view.

## 5  Conclusion

Our contribution proposes a computational model that provides feasibility results for the proposal that perception of form is closely related to perception of motion in the context of natural scenes. We use architectural indoor space as natural scenes that are sampled by eye movement and spatial organization of planes are detected through reconstruction of parallel lines from vanishing points. We demonstrate that the same kind of local computations that must process the stream of such images in a scene are equivalent to the similar local computations for perception of motion from optical flow. These results and the well-established results on cortical neurophysiology of motion naturally suggest that the brain area known as MSTd contains the local circuits for perception of architectural form from reconstruction of vanishing points. Applications include new models for human-like robot motion planning in indoor spaces.